\documentclass[pre,aps,twocolumn]{revtex4}
\usepackage{epsfig}
\usepackage{bm}
\newcommand{\B}[1]{{\bm{#1}}} 
\usepackage[latin1]{inputenc}
\newcommand{\beq}{\begin{equation}}
\newcommand{\eeq}{\end{equation}}
\newcommand{\bea}{\begin{eqnarray}}
\newcommand{\eea}{\end{eqnarray}}
\begin{document}
\title{Dynamical Instabilities of Quasi-static Crack
Propagation Under Thermal Stress\\
Version of \today}
\author{Eran Bouchbinder$^1$, H. George E. Hentschel$^{1,2}$ and Itamar Procaccia$^{1.3}$}
\affiliation{$^1$Dept. of Chemical Physics, The Weizmann Institute
of Science, Rehovot 76100, Israel\\
$^2$Physics Department, Emory University, Atlanta Georgia\\
$^3$Dept. of Physics, The Chinese University of Hong Kong,
Shatin, Hong Kong}
\begin{abstract}
We address the theory of quasi-static crack propagation in a strip of glass
that is pulled from a hot oven towards a cold bath. This problem had been
carefully studied in a number of experiments that offer a wealth of data
to challenge the theory. We improve upon previous theoretical treatments
in a number of ways. First, we offer a technical improvement of the discussion
of the instability towards the creation of a straight crack. This improvement
consists of employing Pad\'e approximants to solve the relevant Weiner-Hopf factorization problem
that is associated with this transition. Next we improve the discussion of the
onset of oscillatory instability towards an undulating crack. We offer a novel
way of considering the problem as a sum of solutions of a finite strip without
a crack and an infinite medium with a crack. This allows us to present a
closed form solution of the stress intensity factors in the vicinity of the 
oscillatory instability. Most importantly we develop a {\em dynamical} 
description of the actual trajectory in the regime of oscillatory crack. 
This theory is based on the dynamical law for crack propagation proposed
by Hodgdon and Sethna. We show that this dynamical law results in a solution
of the actual track trajectory in post critical conditions; we can compute
from first principles the critical value of the control parameters, and the
characteristics of the solution like the wavelength of the
oscillations. We present detailed comparison with experimental measurements
without any free parameter. The comparison appears quite excellent. Lastly we
show that the dynamical law can be translated to an 
equation for the amplitude of the oscillatory crack; this equation
predicts correctly the scaling exponents observed in experiments.
\end{abstract}
\maketitle

\section
{Introduction}
\begin{figure}
\epsfysize=3.2 truecm
\epsfbox{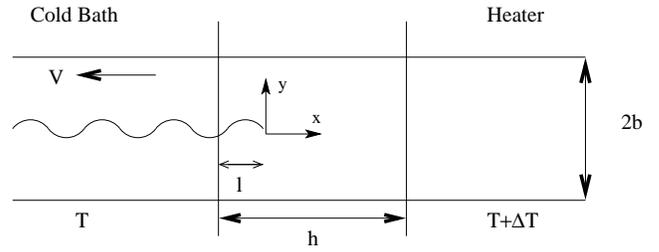}
\caption{Schematic representation of the experiment: a thin
glass plate is pulled at a velocity $v$ away from a heater into a cold bath. The control
parameters are the temperature difference between the oven and the water $\Delta T$, the pulling velocity $v$,
the spatial separation between the thermal baths $h$ and the width of the
plate $2b$.}
\label{sketch}
\end{figure}

In 1993 Yuse and Sano reported a simple experiment on fracture in glass \cite{93YS} that
nevertheless has attracted great attention from the fracture community. The experiment
examined a strip of glass pulled at constant velocity $v$ from an oven into water, cf. Fig. \ref{sketch}.
At small enough velocity nothing happens. A first critical velocity heralds the
onset of a straight crack, whereas exceeding a second critical velocity results in
an oscillatory crack. Finally, at sufficiently high velocities the crack pattern exhibits
multiple fractures and disorder. The reason for the high interest in this relatively simple
experiment is of course that it offers a challenge for the theoretical description of
fracture processes. Being essentially a ``quasi-static" process, as the velocity $v$
is very much smaller than the Rayleigh speed, the fracture process here is free of
many of the complications arising in truly dynamic fracture \cite{FM}. Nevertheless,
in the absence of a microscopic theory of the ``process zone" (how materials
actually break) even the dynamics of
quasi-static crack propagation in brittle materials remains a debatable issue.

The lack of dynamical theory for the cracking process does not hinder the understanding
of the onset of {\em straight} cracks in the above experiment. Indeed, already in the
year following the original experimental observations, Marder has set up the equations describing
the effect of the temperature field
on the elastic theory of the material, and presented a qualitative description of
the onset of straight cracks \cite{94Mar}. From the quantitative point of view this treatment
was lacking, in particular the fracture energy turned out to be strongly
velocity dependent against physical intuition. The tools employed by
Marder did not allow however a correct prediction of the 
characteristics of the oscillatory crack propagation. The
next decisive theoretical step was taken by Adda-Bedia and Pomeau \cite{95A-BP}. These authors
reproduced Marder's results for straight cracks, but also developed a successful criterion
for the secondary instability to oscillatory cracks. They employed the
universal form of the near-tip stress tensor field, i.e.
\begin{equation}
\sigma_{ij}(r,\theta)=\frac{K_I}{\sqrt{2\pi r}}\Sigma^1_{ij}(\theta)+
\frac{K_{II}}{\sqrt{2\pi r}}\Sigma^2_{ij}(\theta) \ .
\label{universalform}
\end{equation}
Here $K_{I}$ and $K_{II}$ are the ``stress intensity factors"
with respect to the opening and shear modes, whereas
$\Sigma^1_{ij}(\theta)$ and $\Sigma^2_{ij}(\theta)$ are universal
angular functions common to all configurations and loading
conditions. Adda-Bedia and Pomeau invoked the  well-known and
extensively used ``principle of local symmetry", which states
that the path taken by a crack in brittle homogeneous isotropic
material is such that the local stress field at the tip of the
crack is of mode I type, annulling $K_{II}$. The considerations
in \cite{95A-BP} led to the conclusion that the appearance of a
negative $K_{II}$ for positive deviations from straight
trajectory (or positive $K_{II}$ for negative deviations) was
tantamount to the onset of the oscillatory instability. Nevertheless
these authors did not offer a careful quantitative comparison against 
the experiments knwon at the time. Their prediction of the fracture
energy and the wavelength of oscillations differed significantly
from the experimental values.

In light of these results, we should explain at this point why do
we feel that further theory is called for. First, we point out
that ``principle of local symmetry" is hardly a dynamical theory.
It can predict an instability, but taken literally would only
agree with a fracture path that has sharp kinks. It cannot be
employed to predict the actual trajectory of a slowly moving
crack when the latter is not straight. Second, since the
theoretical works cited above there have been additional
experimental studies of this very same problem
\cite{95RHP,97YS,98RP}, offering a wealth of data to challenge
the theory, a challenge that had not been picked up by the
theorists. Last, but not least, we feel that we can improve on a
number of technical issues tackled by previous authors; these
will be spelled out in the sequel, hopefully gratifying the
diligent reader as we go along.

From the conceptual point of view we offer a point of departure from previous
treatments by adopting a {\em dynamical} description of the crack development. In this
we follow Hodgdon and Sethna \cite{93HS} who have built
upon the principle of local symmetry, using standard theoretical methods, to reach a
dynamical law for crack propagation which is given by
\begin{eqnarray}
\frac{\partial\B r_{tip}}{\partial t} = v ~ \hat \B t\nonumber\\
\frac{\partial\hat \B t}{\partial t} = -f ~ K_{II}~ \hat \B n
\label{dynamics}
\end{eqnarray}\\
where $\hat \B t$  and  $\hat  \B  n$ are the tangent and the normal to the
crack tip respectively, and $f>0$ is a material parameter that we assume to
be nearly
independent of $\hat \B t$ and $\hat  \B  n$ in
the quasi-static limit.
This law predicts a {\em differentiable} crack path such that $K_{II}$ is reduced.
We will demonstrate that this law of motion provides us with predictions
which are in excellent agreement with the characteristics of the crack trajectory
in the oscillatory regime.
We believe that this is the first context in which Eqs. (\ref{dynamics}) are
compared against a challenging set of experimental data; The comparison appears
quite favorable.

In Sect. \ref{straight} we discuss, facing the danger of being
superfluous, the problem of the primary instability leading to a
starting crack propagation once more. This instability had been
correctly treated in \cite{94Mar,95A-BP}, but we offer a
technical improvement in the handling of the Weiner-Hopf factorization problem
that is the basis of the solution. Using recent mathematical
advances \cite{00Abr} we use Pad\'e approximants to significantly
improve the treatment. Since the results of this improved
treatment are relied upon in our solution of the oscillatory
instability, we present the theory for the straight crack in some
detail. Sect. \ref{osci} introduces the main novel results of our
study in the context of the secondary instability to oscillatory
cracks. Using the dynamical law Eq. (\ref{dynamics}) we show that
as one crosses the second critical value of the parameters the
solution of the equations changes its nature. We solve the
equation near the onset of the oscillatory instability and
calculate the critical values of the control parameters, the
wavelength of oscillations and the material function $f$.
Although our criterion for the oscillatory instability is in
agreement with \cite{95A-BP,96A-BB} (which was based on the
``principle of local symmetry"), we can go considerably further
in describing the actual dynamics in the oscillatory regime. In
particular we present a quantitative comparison with the
experiments. Our handling of the oscillatory instability includes
also a technical improvement on the analysis of \cite{95A-BP};
the latter needed a separate Weiner-Hopf problem for every order
in the amplitude of the perturbation. In our calculation we
derive a new expression for $K_{II}$ to leading order in the
amplitude of the oscillations, an expression that requires a
solution of only one Wiener-Hopf problem. This simplification is
achieved by presenting a new way to decompose the straight crack
problem into a singular and non-singular parts and then using a
classical result of Cotterell and Rice \cite{80CR}. A crucial step
in the calculation is the Wiener-Hopf factorization, for which we
apply the new method of solution based on Pad\'e approximants
\cite{00Abr}. Employing the dynamical law of crack-tip propagation we
calculate the critical exponents for the transition and compare
them with the experimental data. Sect. \ref{conclusion} offers a
summary and concluding remarks.

\section{The straight crack}
\label{straight}

\subsection{Preliminaries}
By varying the experimental control parameters one varies the amount of elastic energy
stored in the glass plate. One can choose various paths in parameter space; in this
work we adopt the scheme of \cite{98RP}, fixing the
values of $\Delta T$, $h$ and $v$. The growth state depends then on the plate's width
$2b$: for small enough values of the width a seeded crack does not grow; For a
width greater than a critical value $L_c$, a crack,
whose tip penetrates a length $\ell$ away from the cold bath, moves at
a velocity $-v$. This crack is stationary in the laboratory frame of
reference and is stable as long as the width is smaller than another critical
value $L_{\rm osc}$. Above this value the crack becomes unstable and
exhibits an oscillatory lateral motion with a well-defined
amplitude and wavelength, still traveling at a velocity
$-v$. As the width is further increased the propagation becomes less and less
regular.

The no crack - straight crack propagation transition is well-understood and the
agreement with the experimental data is favorable \cite{95RHP}. In this case the
propagation is pure mode I and the transition is governed by the following Irwin's relation
\begin{equation}
\frac{K_I^2} {E} = \Gamma
\label{Irwin}
\end{equation}\\
where $\Gamma$ is the fracture energy which is a material property and $E$ is
Young modulus. In
this section we address again this problem and cite the formal solution. The
merit of our treatment will be in
providing  a detailed scheme for performing the Wiener-Hopf factorization in a novel way.
\subsection{The formulation of the problem}

Imagine the glass plate as in Fig. \ref{sketch} with a straight crack penetrating into the glass from
the water side. We choose a coordinate system such that $x=0$ is at the crack tip, (marking the
water level at $x=-\ell$ where $\ell$ is the penetration depth of the straight
crack). The $y$ coordinate spans the interval $[-b,b]$. The
condition for mechanical equilibrium under plane strain conditions caused by a nonuniform temperature field reads
\begin{equation}
\nabla^2\nabla^2\chi\left(x,y\right) = -E\alpha_T\nabla^2T\left(x\right) \ ,
\label{basiceq}
\end{equation}
where $\alpha_T$ is the thermal expansion coefficient and $\chi$ is the
Airy potential which is related to the stress tensor by
\begin{equation}
\sigma_{xx} = \frac{\partial^2\chi} {\partial y^2}, \hspace{1cm}
\sigma_{yy} = \frac{\partial^2\chi} {\partial x^2}, \hspace{1cm}
\sigma_{xy} = -\frac{\partial^2\chi} {\partial x \partial y} \ .
\end{equation}
Using the symmetry of the problem we state the boundary conditions as follows:
\begin{equation}
\sigma_{yy}\left(x,\pm b\right) = \sigma_{xy}\left(x,\pm b\right) = \sigma_{xy}\left(x,0\right) = 0 \\
\end{equation}
\begin{equation}
\sigma_{yy}\left(x,0\right) = 0 ~ \mbox{for $x\leq 0$}, \quad
u_{y}\left(x,0\right)= 0 ~ \mbox{for $x\geq 0
$}\ .
\label{BC}
\end{equation}
Fourier transforming Eq. (\ref{basiceq}) in the $x$ direction and focusing on the upper half
plate one obtains \cite{94Mar} the following Wiener-Hopf equation \cite{58Nob}
\begin{equation}
\hat{\sigma}_{yy}\left(k,0\right) = -F\left(k\right)\hat{u}_y\left(k,0\right) +
D_\ell\left(k\right)
\label{WH}
\end{equation}\\
with\\
\begin{equation}
F\left(k\right) = E k \frac{sinh^2(kb)-kb^2} {sinh(2kb)+2kb}
\end{equation}
\begin{equation}
D_\ell\left(k\right) = 2 E\alpha_T\hat{T}_\ell\left(k\right)
\frac{\left[1-cosh(kb)\right]\left[sinh(kb)-kb\right]} {sinh(2kb)+2kb}
\end{equation}
where one still has to obey the boundary conditions of Eq. (\ref{BC}). Note that
the subscript $\ell$ denotes the transformation $x\rightarrow x+\ell$ in the
temperature field such that the origin of the coordinates system is at the tip
of the crack. For convenience, from now on, we rescale all lengths in the problem by the half
width $b$.

Writing $F\left(k\right) = \frac{F^-\left(k\right)} {F^+\left(k\right)}$,  where
$F^-\left(k\right)$ has neither zeros nor singularities for
$Im\left(k\right)<0$,  and $
F^+\left(k\right)$ has neither zeros nor singularities for
$Im\left(k\right)>0$,  the Wiener-Hopf method \cite{58Nob} results in
\begin{equation}
\hat{u}_y\left(k,0\right) \!= \!\frac{1}{F^-\left(k\right)}\!\!\int_{-\infty}^0
\!\!\!d\tilde{x}\left[\!\!\int_{-\infty}^\infty\frac{d\tilde{k}} {2
\pi} D_\ell(\tilde{k}) F^+(\tilde{k}) e^{-i\tilde{k} \tilde{x}}\right]\!\! e^{ik\tilde{x}} \ .
\end{equation}
From this solution one can extract \cite{95A-BP} the mode I stress intensity factor introduced in
Eq. (\ref{universalform}),
\begin{equation}
K_{I}\left(\Delta T,v,b,h\right) = \int_{-\infty}^\infty\frac{dk} {2
\pi} D_\ell(k) F^+(k)
\label{SIFmodeI}
\end{equation}
Note that if the fracture energy $\Gamma$ is known then the mode I stress intensity factor
characterizes completely the no crack - straight crack transition. $K_I$ is a
positive quantity and is different from zero, as a function of $\ell$, only on
a scale of the order of $b$.
To calculate this quantity we need first to solve for the
temperature field and second to provide
a method for accomplishing the Wiener-Hopf factorization.
\subsection{The temperature field}
\label{Tfield}
The nonuniform temperature field induces the stress field in the elastic plate. In this
subsection we solve for the temperature and study its properties. For simplicity, we set the zero of the
coordinates system at the cooling front level to avoid $\ell$ dependence which is unnecessary
in the present context. For
later calculations we will use the aforementioned transformation to put back the $\ell$
dependence.

In the frame of reference of the plate the temperature field obeys the heat equation\\
\begin{equation}
\frac{\partial T} {\partial t} = D\nabla^2T
\end{equation}\\
with the boundary conditions
\begin{eqnarray}
\vec{\nabla} T\cdot\vec{n} &=& 0\nonumber\\
T\left(x=0\right) &=& T \nonumber\\
T\left(x=h\right)&=&T+\Delta T  .
\end{eqnarray}
Here $D\simeq 0.47mm^2/sec$ is the diffusion coefficient of the glass, h is the distance between the
cold bath and the heater and $\vec{n}$ is the unit vector normal to the boundary of the domain.

This equation can be simplified for the straight crack configuration, for which there is no
$y$ dependence, by looking for a stationary solution in the laboratory frame of reference, of the form $T(x-vt)$.
 This solution obeys the stationary diffusion equation\\
\begin{equation}
\nabla^2T + \frac{1} {d_{th}}\frac{\partial T} {\partial x} = 0 \ ,
\end{equation}\\
where $d_{th}= \frac{D} {v}$ is the thermal diffusion length. The exact solution of this
equation is\\
\begin{equation}
T\left(x\right) = \Delta T \left[\frac{1-e^{-x/d_{th}}} {1-e^{-h/d_{th}}}\theta\left(x\right)
\theta\left(h-x\right)+\theta\left(x-h\right)\right] \ .
\label{tempfield}
\end{equation}

We can identify two distinct regimes, as shown in Fig. 2.
\begin{figure}
\epsfysize=6 truecm
\epsfbox{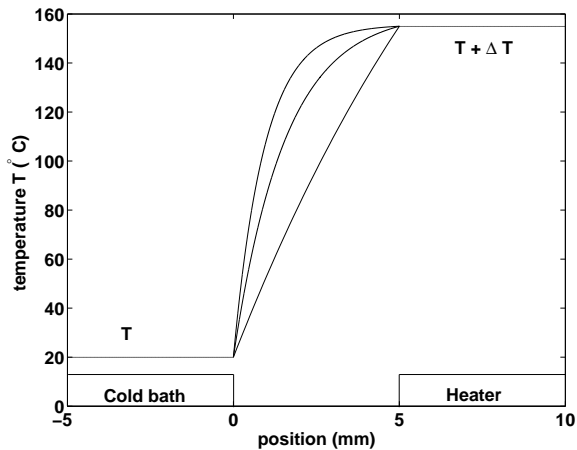}
\caption{The calculated temperature distribution inside the glass
plate. We can identify two distinct regimes, the diffusive regime
at low velocities ($v=0.05mm/sec$ in the lower curve), and the
advective regime at higher velocities ($v=0.3mm/sec$, and
$v=0.5mm/sec$ in the upper two curves). These curves should be
compared with the measured temperature field in \cite{98RP}.}
\label{temperature}
\end{figure}
The first one, at low velocities, is the diffusive regime, in which the
temperature field is controlled by the spatial separation of the thermal baths
h. The second one, at higher  velocities, is the advective regime, in which the
temperature field is controlled by the thermal diffusion length $d_{th}$.
Actually, there is a third regime, at still higher velocities, in which the
temperature field is controlled by the thickness of the plate. Note that we
assumed that the temperature is uniform along this dimension and therefore the breakdown of
this assumption in this regime leads
to a three-dimensional problem, which is outside the scope of our two-dimensional
model.

The temperature field enters the problem through the term
$-E\alpha_T\nabla^2T\left(x\right)$ of the inhomogeneous Bi-Laplace equation. This term is
sensitive only to variations of the temperature gradient, i.e. to the
curvature of the thermal field. In Eq.  (\ref{tempfield}) we considered a finite
spatial separation between the thermal baths, but assumed perfect thermal baths, an
assumption that leads to a discontinuity of the gradient near the baths. This
discontinuity results in incorrect estimates of the stress field; since
at low velocities (see Fig. 2) the only gradient variations are in
these regions, we cannot expect a very good quantitative agreement with the
experiment for low velocities. At higher velocities, there is a significant
curvature inside the glass, so we expect a better quantitative agreement with the
experiment. In \cite{98RP} the temperature field was measured and found to
vary smoothly near the baths due to the finite impedances of the baths. In the
absence of the experimental data of the measured temperature field we will use the
ideal baths approximation.
\subsection{The Wiener-Hopf factorization}

The crucial element in the solution is the Wiener-Hopf factorization of the
known kernel $F\left(k\right)$. Generally it is not possible to find an {\em exact} factorization,
so one tries an approximant $\tilde F(k)\simeq F(k)$ that can be exactly
factorized. This can be made rigorous following a recently proven theorem \cite{00Abr} which
establishes the closeness of the product factors of $\tilde{F}\left(k\right)$ to
those of $F\left(k\right)$ in their region of regularity if
$\tilde{F}\left(k\right)\simeq F\left(k\right)$ for all $k\in{D}$, where $D$ is
the strip of analyticity of $F\left(k\right)$. A commonly followed first step in
finding $\tilde F(k)$ is to examine the behavior of
$F(k)$ near zero and $\pm\infty$\\
\begin{equation}
F(k) \longrightarrow \pm\frac{k} {2} \quad \text{as}~ k\rightarrow\pm \infty
\label{asymp}
\end{equation}
\begin{equation}
F(k) \longrightarrow \frac{k^4}{12} \quad \text{as}~ k\rightarrow \pm 0\\
\end{equation}\\

A standard approach to finding a good factorization is then to seek a function $\phi(k)$
that reproduces the asymptotic behavior of $F(k)$ and to correct it by a ratio of two
polynomials
\begin{equation}
\tilde{F}(k) = \phi(k) \frac{k^4+\alpha
k^2+\beta }
{k^4+\gamma k^2+\beta}
\end{equation}
where for example
\begin{equation}
\phi(k) = \frac{k^4} {\sqrt{4k^6+144}}\ ,
\end{equation}
and $\alpha,\beta,\gamma$ are free parameters that should be chosen as to best fit
$\tilde{F}\left(k\right)$ to $F\left(k\right)$. In principle, one can use higher
order polynomial ratio to achieve greater accuracy. The disadvantage of this
approach is that the positions of the poles and zeros are not well-controlled and
that the convergence behavior of the process is not clear.

In our work we follow a new method developed in \cite{00Abr}. In the heart of this approach lies the use
of Pad\'e approximants. A
$\left[\frac{N} {M}\right]$ approximant of $F\left(k\right)$ is written
as\\
\begin{equation}
\tilde{F}\left(k\right) \simeq  \frac{P_N\left(k\right)} {Q_M\left(k\right)},
\end{equation}\\
where\\
\begin{equation}
P_N\left(k\right) = a_0+a_1k+a_2k^2+ \cdots+a_Nk^N
\end
{equation}
\begin{equation}
Q_M\left(k\right) = 1+b_1k+b_2k^2+ \cdots+b_Mk^M
\end{equation}\\

The coefficients $a_n,b_n$ are determined from the Taylor-series expansion of
$F\left(k\right)$ at any regular point. Let us take the
expansion point to be $k=0$, so
\begin{equation}
F\left(k\right) = \sum_{n=0}^\infty c_nk^n
\end{equation}
where $c_n$ are known. In order to solve for the unknown coefficients one should
set
\begin{equation}
\frac{P_N\left(k\right)} {Q_M\left(k\right)} + \sl O\left(k^{N+M+1}\right)= \sum_{n=0}^\infty c_nk^n
\end{equation}\\
to obtain a set of linear equations \cite{00Abr}. In this method one approximates directly the
factorization and the process is completely algorithmic. Note that since in practice one
uses the truncated series of $F\left(k\right)$ it is not possible to approximate
directly the product factors for arbitrary large $k's$. In order to overcome this
difficulty we should find the asymptotic form of the factorization and use it as a
boundary condition for the Pad\'e approximants.
The asymptotic factorization is found by noticing that the zeros and poles of $F(k)$, which are respectively the solutions of the equations\\
\begin{equation}
\sinh^2(w_n)-w_n^2=0
\label{zeros}
\end{equation}
\begin{equation}
\sinh(2z_n)-2z_n=0
\label{poles}
\end{equation}
have the property that if, for example, $w_n$ is a zero of $F(k)$ then
$\bar{w}_n,\hspace{.15cm} -w_n \hspace{.15cm} and \hspace{.15cm}-\bar{w}_n$ are also zeros. The same holds for the poles.
Therefore, considering the solutions of Eqs.(\ref{zeros}-\ref{poles}) only in the first quadrant,
we obtain
\begin{equation}
F(k)=\frac{k^4}{12}
\frac{\prod^{\infty}_{n=1}(1-\frac{k}{w_n})(1+\frac{k}{w_n})(1-\frac{k}{\bar{w}_n})(1+\frac{k}{\bar{w}_n})}
{\prod^{\infty}_{n=1}(1-\frac{k}{z_n})(1+\frac{k}{z_n})(1-\frac{k}{\bar{z}_n})(1+\frac{k}{\bar{z}_n})} \ .
\end{equation}
From here it follows that
\begin{equation}
F^-(k)=\frac{k^2}{\sqrt{12}}\frac{\prod^{\infty}_{n=1}(1-\frac{k}{w_n})(1+\frac{k}{\bar{w}_n})}
{\prod^{\infty}_{n=1}(1-\frac{k}{z_n})(1+\frac{k}{\bar{z}_n})}=\frac{1}{F^+(-k)} \ .
\end{equation}
Using this relation and the asymptotic relation of Eq.  (\ref{asymp}) we conclude that the asymptotic factorization
is
\begin{equation}
F^+\left(k\right) \rightarrow \sqrt{\frac{2} {-ik}} ,\hspace{.6cm} F^-\left(k\right) \rightarrow
\sqrt{\frac{ik} {2}}
\end{equation}
\begin{figure}
\epsfysize=4.8 truecm
\epsfbox{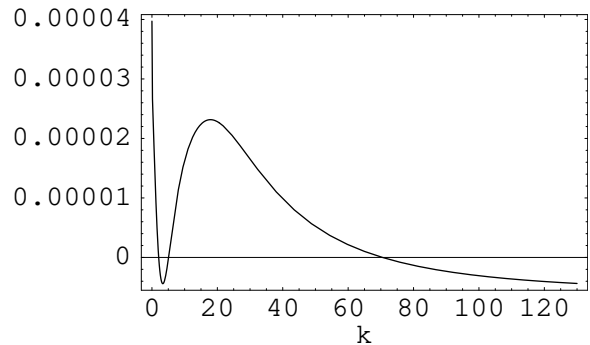}
\caption{The relative error, $\frac{F^-(k)/F^+(k)-F(k)}{F(k)}$,
as a function of $k$.}
\label{pade}
\end{figure}

We should choose the Pad\'e approximants to match these asymptotic forms.
This is achieved by squaring the original kernel
$F(k)$, to obtain an even function of $k$ that behaves as $k^2$ as
$\left|k\right|\rightarrow \infty$. Hence, we can derive an $\left[\frac{N+2} {N}\right]$
Pad\'e approximation
\begin{equation}
F^2(k)\simeq\frac{P_{N+2}\left(k\right)} {Q_N\left(k\right)}
\end{equation}\\
with N even. This approximation contains $N+2$ zeros and $N$ poles which
become, after taking the square-root, $N-1$ branch points in the upper half-plane and
$N-1$ branch points in the lower half-plane. The relative error,
$\frac{F^-(k)/F^+(k)-F(k)}{F(k)}$, for $N=28$ is shown in Fig. 3.
The asymptotic matching of $F^+(k)$ to $\sqrt{\frac{2} {-ik}}$ is shown in
Fig. 4.
\begin{figure}
\epsfysize=3.5 truecm 
\epsfbox{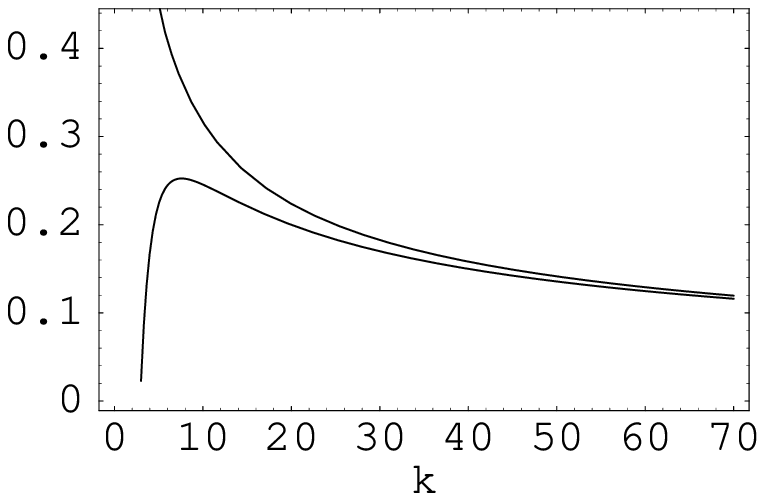}
\epsfysize=3.5 truecm 
\epsfbox{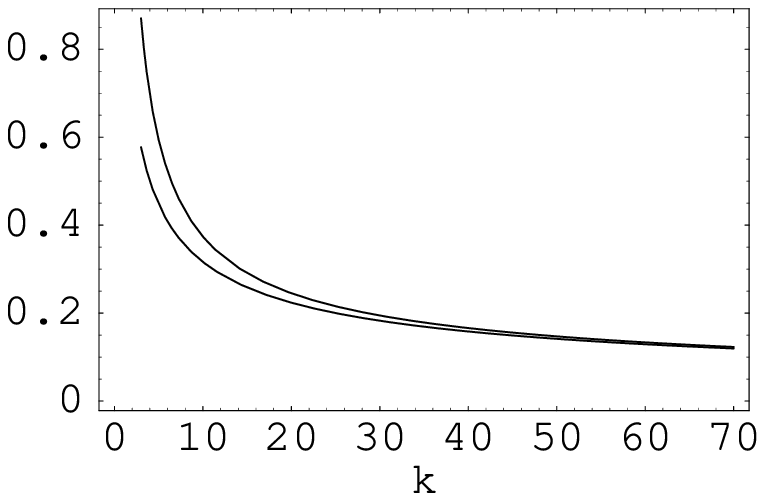}
\caption{The real and imaginary parts of $F^+(k)$ compared to the
real and imaginary parts of $\sqrt{\frac{2} {-ik}}$,  as a
function of $k$.}
\label{pade1}
\end{figure}

\subsection{The determination of the fracture energy $\Gamma$}
\label{Gamma}

In Sect. \ref{osci} we develop a theory for the oscillatory instability and 
compare it with the experimental results. A crucial parameter in that theory,
as here, is the fracture energy $\Gamma$. We can extract the fracture energy from the experimental
threshold for propagation, cf Eq. (\ref{Irwin}). The relevant measurement was reported in 
\cite{95RHP} in which a linear elastic model identical to ours was used to extract the fracture energy.
$\Gamma$ was chosen to best fit the experimental data of the onset of the straight crack
propagation. It was found that $\Gamma$ depends weakly on the velocity when 
the idealized thermal profile was employed; $\Gamma$ turned out velocity independent for 
the actual thermal profile measured in the experiment. Since
we do not have the experimental data for the  thermal profile we will use, for consistency, a typical value of the
former, i.e. $\Gamma\simeq 3.75~J/m^2$. This value should be compared to Fig. 5 in Ref. \cite{95RHP}.
\section{The straight to oscillatory crack transition}
\label{osci}

Solving a dynamic fracture problem in the quasi-static limit consists of (i) solving the equilibrium
equations for the stress field together with a given set of boundary conditions at the sample boundaries
and on the (a-priori unknown and evolving) crack boundary and (ii) employing a
dynamical principle to evolve the crack.  A proper solution determines the correct shape of
the crack as
a function of time. Clearly, the predictions of the employed crack growth law should be consistent
with the experimental observations. In this section we use the dynamical law in Eq.  (\ref{dynamics}) to study the
straight to oscillatory crack transition.

Rewriting the tangential and normal unit vectors at the
tip of the crack in terms of the angle $\theta$ that the tangential unit vector
makes with the $x$-axis we obtain
\begin{eqnarray}
\hat \B t=\cos{\theta}\hspace{.1cm}\bf\hat{x}+\sin{\theta}\hspace{.1cm}\bf\hat{y}\nonumber\\
\hat  \B  n=-\sin{\theta}\hspace{.1cm}\bf\hat{x}+\cos{\theta}\hspace{.1cm}\bf\hat{y}
\end{eqnarray}
which, upon substitution into Eq.  (\ref{dynamics}), leads to
\begin{equation}
\label{GrowthLaw}
\frac{\partial\theta}{\partial t}=-f K_{II}
\end{equation}

This equation predicts that as long as $K_{II}= 0$, the crack will propagate in a
straight line. Nevertheless, in any real material there exist intrinsic
instabilities, due to imperfections of the material and the loading conditions, which produce a small random
$K_{II}\neq 0$.
We are now facing two distinct questions: Under what conditions, in terms of the
control parameters $\Delta T$, $h$, $v$ and $2b$, the straight crack propagation becomes
unstable? Once the straight crack propagation becomes unstable, what is the
stable stationary path that it follows?

The criterion of stability arises naturally from the dynamical equation. If
$\theta$ and $K_{II}$ have the same sign, with $f>0$, then $\frac{\partial\theta}{\partial
t}$ has the opposite sign and $|\theta|$ decreases, which means that a small
perturbation decays. By the same argument, for $\theta$ and $K_{II}$ having the
opposite sign a small perturbation grows. This criterion is identical to the one
suggested in Refs.  \cite{95A-BP,96A-BB}.

The question of the future evolution of the crack, once the instability threshold
was reached, should be answered by solving the dynamical equation just above the
onset of the instability. Guided by the experimental observation that the shape of the crack just above the
onset of the oscillatory instability is a pure
sine function, we introduce a smooth
deviation from straight crack path
\begin{equation}
\label{perturbation}
y(x,t)\simeq A \sin[w(x+vt)] + {\cal O}(A^3) \hspace{.3cm} \mbox{for $x\leq 0$}
\end{equation}
This assumption serves two roles: first, it represents a single mode component,
corresponding to a wavenumber $w$, in the linear decomposition of a small random
perturbation on top of the straight crack and will enable us, for $t=0$, to
analyze its
stability; Second, it is an ansatz for the solution of the dynamical equation
just above the onset of instability. Note that $y(x,t)$ satisfies
$|y(x,t)|\ll1$ and $|y'(x,t)|\ll1$. 
 
In this section we will study the stability
of the straight crack as well as the time evolution of the crack after the onset
of instability by analyzing the dynamical model of Eq. (\ref{GrowthLaw}). The
stability is studied by applying the stability criterion derived above for which
an expression of $K_{II}$ to leading order in the perturbation amplitude in needed. We
derive this expression by introducing an auxiliary problem, the so-called
decomposition problem, whose effective solution enables one to significantly
simplify the derivation. The critical point is calculated by solving the set of
equations governing the transition. The time evolution of the crack just above
the critical point is studied by directly solving the equations of motion for
the tip of the crack.
\subsection{The decomposition problem}

In order to study the stability of the straight crack to small
perturbations we arbitrarily choose $t=0$ in Eq.
(\ref{perturbation}). This choice will be shown later to be
legitimate. We are interested first in finding an expression for
$K_{II}$ to leading order in the amplitude of the perturbation.
We begin by formulating an auxiliary problem. The presence of a
crack, which is usually modeled as a mathematical branch cut,
introduces the famous square-root singularity of the stress field
near the tip of the crack. We want to represent our problem as a
sum of two parts. The first part contains no singularity,
implying that it is crack free, but it includes the geometry of
the problem and the thermal field. The second part contains the
singularity, implying that there is a crack with a given load on
it, but the domain is infinite. The non-singular part is chosen
such that it reproduces the required boundary conditions on the
plate's edges and on the crack. 

Once we obtain the load on the
semi-infinite straight crack in an infinite medium we can apply
the classical result of Cotterell and Rice for slightly curved
cracks\cite{80CR}. The mathematical formulation of this
decomposition problem leads to a set of integro-differential
equations whose complexity may cast doubt on the usefulness of the
whole procedure \cite{94SSN}. In what follows we will show how to
avoid these mathematical difficulties and effectively solve the
problem.

To see that the solution is almost at hand, suppose for a moment
that we succeeded to solve the problem in this way for a straight
crack and for a given set of the control parameters. The load on
a straight crack in {\em an infinite domain} , which is a 
fictitious quantity, must be a pure mode I
load by symmetry. We denote it as $\sigma^{\rm f}
_{yy}(x,y=0;\ell)$, where we marked explicitly the parametric
dependence on $\ell$.  The mode I stress intensity factor is given
by \cite{53Mus}
\begin{equation}
K_I(\ell)=\sqrt{\frac{2}{\pi}}\int_{-\infty}^0\frac{dx
\sigma^{\rm f} _{yy}(x,y=0;\ell)}{\sqrt{-x}} \label{SIFmush}
\end{equation}
Introduce now the $x$-Fourier transform of $\sigma^{\rm f}
_{yy}(x,y=0;\ell)$, denoted as $\hat{\sigma}^{\rm f}
_{yy}(k,y=0;\ell)$. With this object in mind we rewrite Eq. (\ref{SIFmush}) as
\begin{equation}
K_I(\ell)=\int_{-\infty}^\infty\frac{dk} {2 \pi}
\hat{\sigma}^{\rm f}
_{yy}(k,y=0;\ell)[\sqrt{\frac{2}{\pi}}\int_{-\infty}^0\frac{dx
 e^{-ikx}}{\sqrt{-x}}] \ . \label{yofi}
\end{equation}
On the other hand, we have calculated the same quantity using the
Wiener-Hopf technique (see Eq. \ref{SIFmodeI}).
\begin{equation}
K_{I}\left(\ell\right) = \int_{-\infty}^\infty\frac{dk} {2
\pi} D_0(k) e^{-ik\ell} F^+(k)
\label{SIFmodeI1}
\end{equation}
These two expressions for $K_I(\ell)$, though derived through
completely different mathematical procedures, should be identical
functions of $\ell$. The $\ell$ dependence of the second
expression is given by the phase factor $e^{-ik\ell}$ which
immediately implies that $\hat\sigma^{\rm f}
_{yy}(k,y=0;\ell)=\tilde\sigma_{yy}(k,y=0)e^{-ik\ell}$. We
conclude, by the uniqueness of the Fourier transform, that
\begin{equation}
\tilde{\sigma}_{yy}(k,y=0)=\frac{D_0(k)F^+(k)}{[\sqrt{\frac{2}{\pi}}\int_{-\infty}^0\frac{dx
 e^{-ikx}}{\sqrt{-x}}]}=\frac{D_0(k)F^+(k)}{\sqrt{\frac{2} {-ik}}}
\end{equation}
which effectively solves the auxiliary problem. 

Hence, we have
shown how one can use the Wiener-Hopf solution for a traction-free 
straight crack in a finite configuration in order
to find the effective load on a straight crack in an infinite configuration
via the solution of the decomposition problem. We reiterate that
this load is a fictitious tension on a crack in an infinite domain corresponding to
a traction-free situation in a finite domain.
The solution of this auxiliary problem will enable us later on to
use the powerful tool of the complex potential method that is
most suitable for an infinite domain problems. Other theoretical
treatments that were unable to solve this problem led to
incorrect predictions \cite{94SSN}.
\subsection{The critical point}

The calculation now is straightforward. Let us select a local coordinate system
$\{r,\theta\}$ at every point on the crack, with $r$ being the distance from the
point and $\theta$ the angle, starting with $\theta=0$ for the tangent. In such coordinates
the normal opening stress $T_n(x,y(x,t))\equiv\sigma_{\theta\theta}(x,y(x,t))$ and
the tangential shearing stress $T_t(x,y(x,t))\equiv\sigma_{r\theta}(x,y(x,t))$. Using the load $\sigma^{\rm
f}_{yy}$ we can find these stress components for any small deviation $y(x,t)$ from the straight crack.
Applying these loads to the classical result of Cotterell and Rice \cite{80CR},
with $t=0$, we obtain the following expressions for $K_I$ and $K_{II}$ to leading order in $A$ (see
Appendix A)
\begin{widetext}
\begin{eqnarray}
&&K_I=\sqrt{\frac{2}{\pi}}\int_{-\infty}^0\frac{dx\hspace{.1cm}
\hat\sigma^{\rm f}_{yy}(k,y=0;\ell)}{\sqrt{-x}}+{\cal O}(A^2)\nonumber\\
&&K_{II}(A,w,t=0)=
-\sqrt{\frac{2}{\pi}}\int_{-\infty}^0\frac{dx\hspace{.1cm}
[A~\sin(wx)~\hat\sigma^{\rm f}_{yy}(k,y=0;\ell)]}{(-x)^{3/2}}
+\frac{1}{2}~ A~w\hspace{.1cm}K_I+{\cal O}(A^3)
\label{slightlycurved}
\end{eqnarray}
\end{widetext}
This result shows that only $K_{II}$ is changed to first order in the amplitude of the
perturbation. 
The expression for $K_{II}$ has the general form derived in \cite{96A-BB}. It is the
sum of two competitive terms which the authors of \cite{96A-BB} refer to as a ``physical" shear stress which is
a destabilizing term (the first term), and as a ``geometric" shear stress, which is a
stabilizing term (the second term). We expect to find a range of the control parameters for which for every $w\neq0$
the second term dominates the first, leading to $K_{II}>0$, which implies a stable straight crack propagation.
Thus, our stability criterion states that the transition between straight and oscillatory crack propagation 
occurs when there exist $A,w\neq0$ such that $K_{II}(A,w,t=0)=0$. One concludes that the straight crack to
oscillatory crack transition is governed by the following set of equations
\begin{eqnarray}
\frac{K_I^2(b,\ell)+K_{II}^2(b,\ell,w)}{E}&\simeq& \frac{K_I^2(b,\ell)}{E} =\Gamma\nonumber\\
K_{II}(b,\ell,w)&=&0\nonumber\\
\frac{\partial K_{II}(b,\ell,w)}{\partial w}&=&0
\end{eqnarray}
where we made explicit the dependence on $b$, $\ell$ and $w$.

The first equation is the Irwin's relation which expresses the energy balance between the elastic energy flow
to the tip of the crack and the fracture energy needed to create a new crack
surface. This fracture energy is a parameter of our model; in previous applications this
parameter was optimized for agreement with experiments \cite{95RHP}. We cannot afford such
luxury since we have determined already the parameter in Sec. \ref{Gamma}. Therefore 
in our comparison with experiments the theory is truly challenged, and the agreement will
be shown to be very satisfactory. 

The second and third equations express the stability threshold. In order to characterize quantitatively this transition we
adopt the experimental scheme of Ref. \cite{98RP}, in which $\Delta T$ and $h$ are
kept fixed and for a given $v$ the critical width for the the onset of oscillations, $L_{\rm osc}$,
is found. Fig. 5 shows a typical situation in which $K_{II}/A$ is plotted as a
function of $w$ for different values of the width $2b$ at constant $\Delta T$,
$v$ and $h$. It is shown that as one increases the width a solution for the
equation $K_{II}(A,w,t=0)=0$ appears.
\begin{figure}
\epsfysize=4.9 truecm
\epsfbox{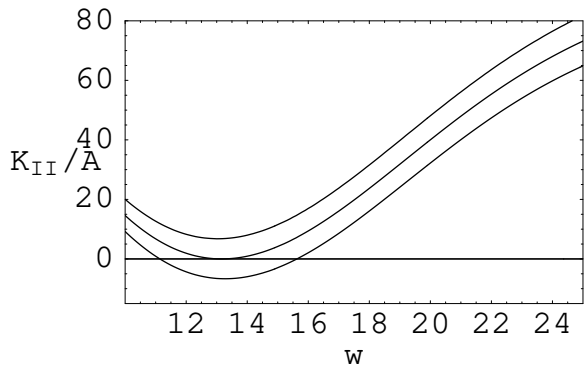}
\caption{$K_{II}/A$ versus the dimensionless wavenumber $w$. For
fixed $\Delta T$, $v$ and $h$, the curves from top down show the
increasing of the stored elastic energy via the increasing of the
width of the plate. It is clear that there is a critical width
for which $K_{II}(w)=0$.}
\label{transition}
\end{figure}

We solved the above set of equations graphically
by the following procedure. We fixed $\Delta T=135^\circ C$ and $h=5mm$ and for
each velocity we changed $b$ until we converged to the solution. Note that first
one has to solve the first equation for $\ell$. Fig. 6 shows the critical width for
oscillations $L_{\rm osc}=2b_{\rm osc}$ as a function of the driving velocity $v$. The
experimental data reported in
Ref \cite{98RP} have been added for comparison.
\begin{figure}
\epsfysize=5 truecm
\epsfbox{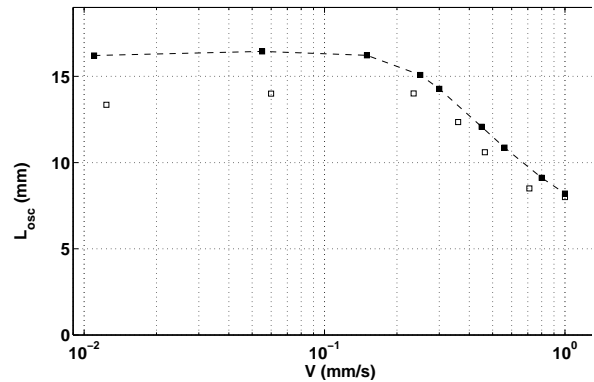}
\label{phasediagram}
\caption{The
critical width for oscillations $L_{\rm osc}=2b_{\rm osc}$ versus the
driving velocity $v$ for $\Delta T=135^\circ C$ and $h=5mm$. We
didn't calculated these quantities for still higher velocities
since this regime is controlled by three-dimensional effects
which are outside the scope of our theory. The theoretical values
are connected by the dashed line that was added as a guide for
the eye. The experimental values that are represented by the
unfilled squares were extracted from Fig. 15 in Ref. \cite{98RP}.
The deviation of $L_{\rm osc}$ from the measured data is due to its
high sensitivity to the fine details of the temperature field,
which we approximated, in the absence of the measured one, using
the ideal baths assumption. It is clear, as predicted in Sec.
\ref{Tfield}, that the agreement with the experiment is much better
within the advective regime than within the diffusive regime.}
\end{figure}

The deviation of $L_{\rm osc}$ from the measured data is due to its high sensitivity
to the fine details of the temperature field, which we approximated, in the
absence of the measured one, using the ideal baths assumption. Nevertheless, as
predicted in Sec. \ref{Tfield}, the agreement with the experiment is much better
within the advective regime, in which the temperature field is controlled by
$d_{th}$, than within the diffusive regime, in which the temperature field is
controlled by $h$.
Our solution here yields also the wavenumber of the unstable mode. In
\cite{96A-BB} this wavenumber was identified with the wavenumber of 
the actual trajectory in the post-critical conditions without further 
discussion. We find this unsatisfactory;  whether
or not this wavenumber will be observed in the actual crack trajectory 
also in post-critical conditions depends on the dynamics. If the ``fastest
growing mode" becomes stabilized by the nonlinear terms, then this wavenumber
would be observed. To assess this issue we must turn next to the weakly non-linear
theory in the post-critical regime. 
\subsection{The post-critical time evolution of the crack}

Crack propagation laws used in the literature so far were unable to predict analytically 
non-trivial trajectories left behind a crack tip. The set of well-controlled experiments described here
offers a challenge to any dynamical law, especially near the critical point where the crack
exhibits a lateral oscillatory motion with a well-defined wavelength and amplitude. In this subsection
we will show that the adopted dynamical law meets that challenge.
The stationary stable path of the crack just above the onset of the oscillatory
instability is determined by the solution of the dynamical equation near the
transition. Noting that under our assumptions $\theta\simeq y'(0,t)$ we obtain
\begin{equation}
\frac{\partial y'(0,t)}{\partial t}=-f K_{II}(A,w,t)+{\cal O}(A^3)
\label{linearDynamics}
\end{equation}
Deriving the time dependent expression for $K_{II}$ (see Appendix
A) and substituting our ansatz (\ref{perturbation}), Eq. (\ref{linearDynamics}) becomes
\begin{widetext}
\begin{equation}
\frac{Aw^2v}{f}\sin(wvt)=
 \sqrt{\frac{2}{\pi}}\int_{-\infty}^0\frac{dx\hspace{.1cm}
\{2A\sin[w(x+vt)]\sigma^{\rm f}_{yy}(x,y=0;\ell)\}'-A\sin(wvt)\sigma^{'\rm f}_{yy}(x,y=0;\ell)
}{\sqrt{-x}}\nonumber\\
+\frac{1}{2}\hspace{.1cm}Aw\cos(wvt)\hspace{.1cm}K_I
\label{shapeEq}
\end{equation}
\end{widetext}

This equation has a trivial solution, i.e. $A=0$, which is the straight crack. An explicit calculation with
$A\neq 0$ determines that the RHS of this equation is a pure sine function.  Thus our ansatz (\ref{perturbation})
can be an actual solution only if we can choose the control parameters such as to set the phase of the sine
function to zero at $t=0$ (cf. the LHS). We see that this is possible with $K_{II}(A,w,t=0)=0$ which is exactly what
was calculated above in the context of linear stability analysis. Thus if this condition can be
met, (and if $A$ remains small above the critical point, cf. the next subsection) we can indeed identify
the aforementioned wavenumber as the wavenumber of the oscillations in the close vicinity of the
critical point. We conclude that the equation of motion (\ref{GrowthLaw}) is consistent with a pure sinusoidal
trajectory, which is an {\em exact solution of the post-critical} dynamics. This is in good agreement with
the experimental observations. This result also shows that the arbitrary choice $t=0$ in the linear stability analysis
is legitimate since one has to fix the phase only at one time point.

Fig. 7 shows  the wavelength of the oscillations
$\lambda_{\rm osc}=2\pi/w_{\rm osc}$ as a function of the driving velocity $v$. The critical width for
oscillations $L_{\rm osc}=2b_{\rm osc}$, first shown in Fig. 6, was added for completeness, while the
experimental data reported in \cite{98RP} have been added for comparison.
\begin{figure}
\epsfysize=5 truecm
\label{phasediagram1}
\epsfbox{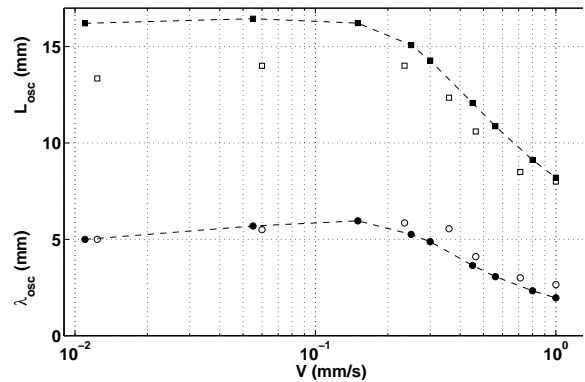}
\caption{The wavelength of the oscillations
$\lambda_{\rm osc}=2\pi/w_{\rm osc}$ versus the driving velocity $v$ for
$\Delta T=135^\circ C$ and $h=5mm$. The theoretical values are connected by the
dashed lines that were added as guides for the eye. The
experimental points that are represented by the unfilled points
were extracted from Fig. 15 in Ref. \cite{98RP}. The results of Fig. 6 were superimposed
for completeness. The
wavelength of oscillations $\lambda_{\rm osc}$ seems far less
sensitive to the fine details of the temperature field than
$L_{\rm osc}$.}
\end{figure}
It is clear that the wavelength of oscillations agrees rather well with the
experimental data, which confirms the assertion in Ref. \cite{98RP} that the
oscillation wavelength $\lambda_{\rm osc}$ seems far less sensitive to the fine details of
the temperature field than $L_{\rm osc}$.

As we have indicated before there are several relevant length scales in the problem. Here we have
calculated a new quantity which has the dimension of length and we wanted to explore its dependence on the
various length scales in the problem. Therefore, we have calculated 
the dimensionless oscillation wavelength $\lambda_{\rm osc}/L_{\rm osc}$
at the threshold of instability as a function of the dimensionless thermal
diffusion length $d_{th}/L_{\rm osc}$. The results are shown in Fig. 8. We have found
that within the advective regime, which corresponds to relatively high velocities,
this function can be well fitted by the linear scaling law
$\lambda_{\rm osc}/L_{\rm osc}\simeq\alpha+\beta d_{th}/L_{\rm osc}$ with $\alpha=0.12$ and
$\beta=2.1$, to be compared with the experimental values of $\alpha=0.15$ and
$\beta=2.5$ \cite{98RP} and the FEM simulation values of $\alpha=0.14$ and
$\beta=2.1$ \cite{95bahr}. This result shows that the scaling between
$\lambda_{\rm osc}$ and $L_{\rm osc}$ is controlled by the thermal diffusion length.
\begin{figure}
\epsfysize=5.5 truecm
\epsfbox{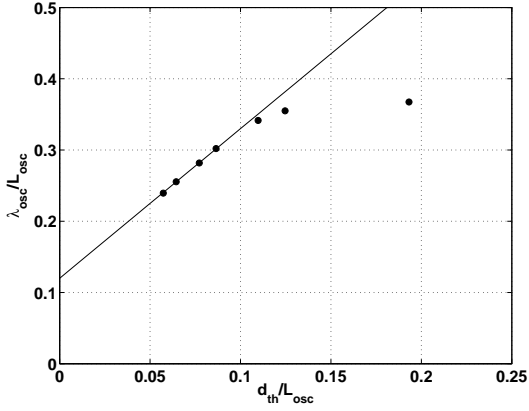} \label{scalindlaw}
\caption{Dimensionless oscillation wavelength
$\lambda_{\rm osc}/L_{\rm osc}$ at the threshold of instability as a
function of the dimensionless thermal diffusion length
$d_{th}/L_{\rm osc}$. It is seen that within the advective regime the
values can be well fitted by a straight line. This result shows
that the scaling between $\lambda_{\rm osc}$ and $L_{\rm osc}$ is
controlled by the thermal diffusion length.}
\end{figure}

Up to now we have dealt with the matching of the phases of both sides of Eq. (\ref{shapeEq}).
The matching of the amplitudes will enable us to calculate the material function $f$. The RHS
of this equation, at the critical point, has the form
\begin{equation}
A E\alpha_T\Delta T\sqrt{b_{\rm osc}}A^*\sin(w_{\rm osc}v_{\rm osc}t)
\end{equation}
where $A^*$ is a dimensionless amplitude to be calculated. By equating the
amplitudes of both sides of Eq. (\ref{shapeEq}) we obtain
\begin{equation}
f=\frac{w_{\rm osc}^2v_{\rm osc}} {E\alpha_T\Delta T\sqrt{b_{\rm osc}}A^*} \ .
\end{equation}
The material function $f$ determines the decay length of perturbations with finite $K_{II}$
back to a pure mode I propagation in the straight crack regime. A typical length
is constructed from $v_{\rm osc}/g$, where
$g\equiv f E\alpha_T\Delta T\sqrt{b_{\rm osc}}$. Fig. 9 shows 
$v_{\rm osc}/g$ (in units of $b_{\rm osc}$) as a function of $v_{\rm osc}$ for
velocities in the advective regime where the theory agrees with the experiment. It is
seen that this length (in units of $b_{\rm osc}$) decreases as the velocity increases. Assuming that this
behavior is not sensitive to the details of the temperature field, it is a
challenge to any theory that will suggest an independent derivation of
$f$.
\begin{figure}
\epsfysize=5.25 truecm
\epsfbox{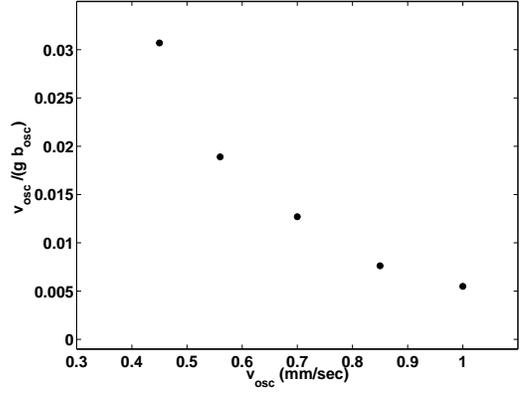}
\label{f_parameter} \caption{$v_{\rm osc}/(g~b_{\rm osc})$ as a function of
$v_{\rm osc}$ for velocities in the advective regime, where $g\equiv
f E\alpha_T\Delta T\sqrt{b_{\rm osc}}$.}
\end{figure}
\subsection{The critical exponents of the amplitude of oscillations}

In the previous subsections we discussed how the solution of the equation of motion changes its
nature from an $A=0$ solution to an $A\ne0$ solution, but since we considered all the relevant
quantities to ${\cal O}(A)$ we could not study the properties of the amplitude
itself. In order to extend our analysis we should introduce a time dependent
amplitude in our ansatz
\begin{equation}
y(x,t)=A(x+vt)\sin[w(x+vt)]
\end{equation}
and write the equation of motion to ${\cal O}(A^3)$. The orientation of the tip 
of the crack is given by
\begin{equation}
\theta(t)=tg^{-1}[y'(x=0,t)]\simeq y'(x=0,t) - \frac{y'^3(x=0,t)}{3}
\end{equation}
which upon substitution of our ansatz becomes
\begin{eqnarray}
\theta(t)&\simeq& w~A(vt)~\cos(wvt)+A'(vt)~\sin(wvt)\\
&-&\frac{1}{3}[w~A(vt)~\cos(wvt)+A'(vt)~\sin(wvt)]^3\nonumber
\end{eqnarray}
where the prime denotes a derivative with respect to the argument $vt$.
Because of the symmetry $A\rightarrow -A$ the next order term in the expansion of 
$K_{II}$ in powers of A is of ${\cal O}(A^3)$. Thus, 
\begin{eqnarray}
K_{II}&\simeq& K^{(1)}_{II}\{A(x+vt)\sin[w(x+vt)]\}\\
&+&K^{(3)}_{II}\{A(x+vt)\sin[w(x+vt)]\}+{\cal O}(A^5) \ , \nonumber
\end{eqnarray}
where $K^{(1)}_{II}\{\cdot\}$ is the functional that was calculated in the Appendix, being of 
${\cal O}(A)$.  $K^{(3)}_{II}\{\cdot\}$
is a functional whose derivation is straightforward but very lengthy; we do 
not present it explicitly here, but note that it yields a term of ${\cal O}(A^3)$.  

In order to proceed we assume that our problem exhibits separation of time scales; the amplitude
changes on a typical time that is much longer than the period of oscillations. Hence, we can
substitute the expressions for $\theta$ and $K_{II}$ into the equation of motion, cf. Eq.
(\ref{GrowthLaw}), and operate on both sides of the equation with the operator
$\frac{wv}{2\pi}\int_0^\frac{2\pi}{wv}\{\cdot\}\sin(wvt)dt$ to obtain
\begin{eqnarray}
&-&vw^2A/2+vw^4A^3/8-vw^2AA'^2/8+vA''/2\nonumber\\
&-&vw^2A^2A''/8-3vA'^2A''/8=\\
&-&\frac{fwv}{2\pi}\int_0^\frac{2\pi}{wv}K^1_{II}\{A(x+vt)\sin[w(x+vt)]\}\sin(wvt)dt\nonumber\\
&-&\frac{fwv}{2\pi}\int_0^\frac{2\pi}{wv}K^3_{II}\{A(x+vt)\sin[w(x+vt)]\}\sin(wvt)dt\nonumber
\end{eqnarray}
This is a highly non-trivial integro-differential equation for 
the time evolution of the amplitude $A$. We expect that after a transient the amplitude 
saturates to a fixed value. Therefore, we set all the derivatives to zero and
the amplitude equation reduces to
\begin{eqnarray}
&-&vw^2A/2+vw^4A^3/8=\\
&-&\frac{fwv}{2\pi}\int_0^\frac{2\pi}{wv}K^{(1)}_{II}\{A\sin[w(x+vt)]\}\sin(wvt)dt\nonumber\\
&-&\frac{fwv}{2\pi}\int_0^\frac{2\pi}{wv}K^{(3)}_{II}\{A\sin[w(x+vt)]\}\sin(wvt)dt\nonumber
\end{eqnarray}

If we consider one of the control parameters slightly above its critical value,
e.g. the velocity, and 
expand all the terms around the critical point we obtain
\begin{eqnarray}
&-&w_{\rm osc}^2A/2-\epsilon_v(w_{\rm osc}^2+2v_{\rm osc}w_{\rm osc}\partial_vw_{\rm osc})A/2\nonumber\\
&+&w_{\rm osc}^4A^3/8=-fAK^{(1)}_{II}\{A\sin[w_{\rm osc}(x+v_{\rm osc}t)]\}/2v_{\rm osc}\nonumber\\
&+&\alpha(v_{\rm osc})\epsilon_v A+\beta(v_{\rm osc}) A^3 \ .\nonumber
\end{eqnarray}
Here $\epsilon_v\equiv\frac{v-v_{\rm osc}}{v_{\rm osc}}$ is the critical parameter,
$\alpha(v_{\rm osc})$ is the coefficient of a critical linear term related to 
$\partial_vK^{(1)}_{II}\{A\sin[w_{\rm osc}(x+v_{\rm osc}t)]\}$,
$\beta(v_{\rm osc})$ is the coefficient of the non-critical cubic term, 
and we neglected terms of ${\cal O}(\epsilon_vA^3)$. 
The linear {\em non-critical} terms cancel out since they are just the equation 
to ${\cal O}(A)$.
Therefore, we are left with
\begin{eqnarray}
0&=&
[(w_{\rm osc}^2+2v_{\rm osc}w_{\rm osc}\partial_vw_{\rm osc})/2
+\alpha(v_{\rm osc})]\epsilon_v A\nonumber\\
&+&[-w_{\rm osc}^4/8+\beta(v_{\rm osc})] A^3\nonumber\\
\end{eqnarray}
where $\partial_vw_{\rm osc}>0$ since as we increase the velocity we increase the
energy flow to the crack tip which requires more crack surface created per unit
time. The instability is tantamount to a positive linear coefficient. We {\em assume}
that the cubic term leads to saturation. It follows that
\begin{equation}
A=\frac{(w_{\rm osc}^2+2v_{\rm osc}w_{\rm osc}\partial_vw_{\rm osc})/2+
\alpha(v_{\rm osc})}{w_{\rm osc}^4/8-\beta(v_{\rm osc})}\epsilon_v^{1/2}
\end{equation}
This result shows that the critical exponent with respect to $\epsilon_v$ is $1/2$,
in agreement with the experimental data\cite{97YS}. A similar calculation can
be presented for the dependence of $A$ on $\epsilon_{\Delta T}\equiv (\Delta T-\Delta T_{\rm osc})/
\Delta T_{\rm osc}$. Such a calculation result with the same exponent 1/2 as 
observed in the experiment.

Two comments are in order. First, we {\em derived} the amplitude equation from
the dynamics of the tip, Eq. (\ref{GrowthLaw}), rather than guess it
as in previous works. More stamina can in principle lead to an actual
calculation of the last term in this equation. Second, we have projected
the full amplitude equation onto its asymmetric part. Projecting onto the
symmetric (cosine) part yields the equation
\begin{eqnarray}
&&vwA'-vw^3A^2A'/2-vwA'^3/4-vwAA'A''/4=\\
&-&\frac{fwv}{2\pi}\int_0^\frac{2\pi}{wv}K^{(1)}_{II}\{A(x+vt)\sin[w(x+vt)]\}\cos(wvt)dt\nonumber\\
&-&\frac{fwv}{2\pi}\int_0^\frac{2\pi}{wv}K^{(3)}_{II}\{A(x+vt)\sin[w(x+vt)]\}\cos(wvt)dt\nonumber \ .
\end{eqnarray}
Discarding again, for the stationary state, all the derivatives, we see that we 
do not gain any new information about the stationary amplitude. On the other
hand we learn that also the last term is a pure sine function, since it has to vanish
at the critical point exactly like the $K^{(1)}_{II}$ term.
\section{Concluding remarks}
\label{conclusion}

The main point of departure of our theory from previous ones is that we employ, in addition to the
two-dimensional linear elasticity part, the dynamical crack-tip propagation law
suggested in Ref. \cite{93HS}. Using this dynamical law we first derived a stability
criterion for the straight crack propagation which is identical to a previously
suggested criterion \cite{95A-BP, 96A-BB}. We then extended the analysis to the
evolution of the crack shape just above the onset of the oscillatory instability
and showed that the dynamical equation has a stationary sinusoidal solution with
a theoretically calculated wavenumber. We presented a quantitative
comparison with the experimental data for a temperature field that is
characterized both by the spatial separation between the thermal baths $h$ and the
thermal diffusion length $d_{th}$. Our results agree rather well with
the experiments \cite{98RP}.

From the conceptual point of view we have offered a successful way to decompose
the problem into a singular and a non-singular parts. This decomposition enabled
us to derive an expression for $K_{II}$ to leading order in the amplitude of the
oscillations that depended only on the factorization of one Wiener-Hopf kernel.
This factorization is done by applying the method of Pad\'e approximants suggested
recently \cite{00Abr}. Finally, we showed how the dynamical tip propagation law
translates to an amplitude equation for the oscillatory solution. This 
equation resulted in calculated critical exponents of the transition, in agreement with the
measured ones. The success of the dynamical theory based on the law of tip 
propagation lends strong support to this law, at least in these quasi-static
conditions. One should stress at this point that the analysis considered
the temperature field as effectively frozen. The oscillatory nature of the
crack has very little to do with the temperature dynamics. This cannot be 
expected to remain valid for larger amplitudes of oscillations
since the boundary condition restrict the temperature level sets to be
normal to the crack. Thus at some point the dynamics of the termperature
field must enter the discussion, potentially leading to new dynamic
instabilities including chaos and disorder.  

In fact, the conclusion of this study appears to be that in the quasi-static conditions
the assumption of small scale yielding holds, making it
sufficient to solve the linear elasticity problem, coupled to a
correct law of motion that dictates how the tip propagates. It would be 
interesting to try to apply this or similar laws to other contexts in which
the quasi-static problem can be solved, but where the absence of an accepted
propagation law has led to a number of possible evolutions \cite{02LP,02BLP}. 
We expect however that in truly dynamical crack propagation new theoretical
concepts need to be developed in order to reach a similar level of calculation
of experimental observations.

\begin{acknowledgments}
We thank Vincent Hakim for proposing the problem to us, suggesting that there
is substantial amount of theory to be done. This work has been supported in part
by the Minerva Foundation, Munich, Germany and by the European Commission under
a TMR grant.
\end{acknowledgments}
\appendix
\section{slightly curved cracks}

\hspace{.6cm}The aim of this appendix is to derive Eq. (\ref{slightlycurved})
and Eq. (\ref{shapeEq}).
Assume that we have a mode I load $\sigma_{yy}(x,0)$ on a semi-infinite crack
whose tip is at $x=0$. Let us first find the normal opening stress $T_n(x,y(x,t))$ and  tangential shearing
stress $T_t(x,y(x,t))$ on any small deviation $y(x,t)$ from straight crack in
terms of $\sigma_{yy}(x,0)$. The cartesian components of the stress
tensor field are related to polar components according to
\begin{eqnarray}
\sigma_{xx}&=&\sigma_{rr}\cos^2{\theta}+\sigma_{\theta\theta}\sin^2{\theta}-\sigma_{r\theta}\sin{2\theta}\nonumber\\
\sigma_{yy}&=&\sigma_{rr}\sin^2{\theta}+\sigma_{\theta\theta}\cos^2{\theta}+\sigma_{r\theta}\sin{2\theta}\nonumber\\
\sigma_{xy}&=&(\sigma_{rr}-\sigma_{\theta\theta})\sin{(2\theta/2)}+\sigma_{r\theta}\cos{2\theta} \ .
\end{eqnarray}
Here $\theta$ is the local angle made by the tangent to the crack and the
$x$-axis. These relations can be expanded to first order in $\theta\simeq y'(x,t)$ and then inverted to yield
\begin{eqnarray}
&&T_n(x,y(x,t))=\sigma_{\theta\theta}(x,y(x,t))\\&&
=\sigma_{yy}(x,y(x,t))-2y'(x,t)\sigma_{xy}(x,y(x,t))\nonumber\\~\nonumber\\
&&T_t(x,y(x,t))=\sigma_{r\theta}(x,y(x,t))=
\sigma_{xy}(x,y(x,t))\nonumber\\&&+y'(x,t)[\sigma_{yy}(x,y(x,t))-\sigma_{xx}(x,y(x,t))] \ . \nonumber
\end{eqnarray}
Expanding the cartesian components to first order in $y(x,t)$
\begin{eqnarray}
\sigma_{yy}(x,y(x,t))=\sigma_{yy}(x,0)+\partial_y\sigma_{yy}(x,0)\hspace{.1cm}y(x,t)\nonumber\\
\sigma_{xy}(x,y(x,t))=\sigma_{xy}(x,0)+\partial_y\sigma_{xy}(x,0)\hspace{.1cm}y(x,t)
\end{eqnarray}
and using the relation
\begin{equation}
\partial_y\sigma_{xy}(x,0)=\partial_y
(-\partial_x\partial_y\chi)=
-\partial_x\partial_y\partial_y\chi=
-\sigma'_{xx}(x,0)
\end{equation}
we end up with
\begin{eqnarray}
\label{loading}
T_n(x,y(x,t))&=&\sigma_{yy}(x,0)\nonumber\\~\nonumber\\
T_t(x,y(x,t))&=&y'(x,t)(\sigma_{yy}(x,0)-\sigma_{xx}(x,0))\nonumber\\
&-&y(x,t)\sigma'_{xx}(x,0)
\end{eqnarray}
where we used the symmetry of the problem to set $\sigma_{xy}(x,0)=0$; $\sigma_{xx}(x,0)$
can be calculated from the knowledge of the boundary condition $\sigma_{yy}(x,0)$.

The problem we should solve now is formulated as follows; Given the following crack
configuration and loading conditions:(i) a semi-infinite crack
whose shape is described by a small deviation $y(x,t)$ from a straight crack configuration in an infinite
two-dimensional domain (ii) a normal opening load $T_n(x,y(x,t))$ and a shear load
$T_t(x,y(x,t))$ at any point on the crack, what are the mixed mode stress intensity factors?

 A version of this problem was treated completely by Cotterell and Rice\cite{80CR}. They have
found that the stress intensity factors for a finite slightly curved crack extending from $-a$
to $a$, where the deviation vanishes at both tips, are given by
\begin{equation}
K_I-iK_{II}=\frac{1}{\sqrt{\pi a}}\int_{-a}^a dx\hspace{.1cm}
[q_I(x)-iq_{II}(x)]\sqrt{\frac{a+x}{a-x}}
\label{CR1}
\end{equation}
where $q_I(x)$ and $q_{II}(x)$ were derived explicitly in \cite{80CR}.
In order to adapt this result to our semi-infinite configuration we should fix one tip of
the crack to $x=0$, take the limit where the other tip goes to $-\infty$ and finally
shift the origin,
\begin{equation}
\tilde{y}(x,t)=y(x,t)-y(0,t) \ ,
\end{equation}
such that the deviation vanishes as the tip of the crack at any time.
This adaptation yields
\begin{equation}
K_I-iK_{II}=\sqrt{\frac{2}{\pi}}\int_{-\infty}^0\frac{dx
[q_I(x)-iq_{II}(x)]}{\sqrt{-x}}
\label{CR}
\end{equation}
where
\begin{eqnarray}
&&q_I=T_n-\frac{3}{2}\tilde{y}'(0,t)T_t+\tilde{y}(x,t)T'_t+2\tilde{y}(x,t)T_t\nonumber\\
&&q_{II}=T_t+\tilde{y}(x,t)T'_n+\frac{1}{2}\tilde{y}'(0,t)T_n
\end{eqnarray}
Using the derived loading conditions of Eq. (\ref{loading}) and applying these 
results to our ansatz for $y(x,t)$ we obtain

\begin{eqnarray}
&&q_I=\sigma_{yy}(x,0)+{\cal O}(A^2)\\~\nonumber\\
&&q_{II}=\{A
\sin[w(x+vt)](\sigma_{yy}(x,0)-\sigma_{xx}(x,0))\}'\nonumber\\
&&-A\sin(wvt)\sigma'_{yy}(x,\!0)\!+\!\frac{1}{2} Aw\cos(wvt) \sigma_{yy}(x,\!0)\!+\!{\cal O}(A^3) \ . \nonumber
\end{eqnarray}

The last step is to relate $\sigma_{xx}(x,0)$
to the boundary condition $\sigma_{yy}(x,0)$. According to the complex potentials method \cite{53Mus}
we have for a semi-infinite straight crack
\begin{eqnarray}
\sigma_{yy}+\sigma_{xx}=4Re[\Phi(z)]
\end{eqnarray}
with
\begin{equation}
\Phi(z)=\frac{1}{2 \pi \sqrt{z}}\int_{-\infty}^0\frac{dx
[\sigma_{yy}(x,0)-i\sigma_{xy}(x,0)]\sqrt{-x}}{z-x}
\end{equation}
Since in our case there is no shear loading on the straight crack we obtain
\begin{eqnarray}
q_I&=&\sigma_{yy}(x,0)+{\cal O}(A^2)\\~\nonumber\\
q_{II}&=&\{2A
\sin[w(x+vt)]\sigma_{yy}(x,\!0)\}'\!-\!A\sin(wvt)\sigma'_{yy}(x,\!0)\nonumber\\
&+&\frac{1}{2}~Aw\cos(wvt)~\sigma_{yy}(x,0)+{\cal O}(A^3)\nonumber
\end{eqnarray}
which becomes, upon substitution into Eq. (\ref{CR}), the RHS of Eq.
(\ref{shapeEq}) and by setting $
t=0$ and integrating by parts gives Eqs. (\ref{slightlycurved}).


\begin{thebibliography}{99}

\bibitem{93YS}
A. Yuse and M. Sano, Nature (London) {\bf 362}, 329 (1993).

\bibitem{FM}
J. Fineberg and M. Marder, Phys. Rep. {\bf 313}, 1 (1999).

\bibitem{94Mar}
M. Marder, Phys. Rev. Lett. {\bf 49}, R51 (1994).

\bibitem{95A-BP}
M. Adda-Bedia and Y. Pomeau, Phys. Rev. E \bf{52}\rm, 4105 (1995).

\bibitem{96A-BB}
M. Adda-Bedia and M. Ben Amar, Phys. Rev. Lett. \bf{76}\rm, 1497 (1996).

  \bibitem{95RHP}
O. Ronsin, F. Heslot and B. Perrin, Phys. Rev. Lett. {\bf 75}, 2352 (1995).

\bibitem{97YS}
A. Yuse and M. Sano, Physica D {\bf 108}, 365 (1997).

\bibitem{98RP}
O. Ronsin and B. Perrin, Phys. Rev. E {\bf 58}, 7878 (1998).

\bibitem{93HS}
J. Hodgdon and J. Sethna, Phys. Rev. B {\bf 47}, 4831 (1993).

\bibitem{00Abr}
I. D. Abrahams, IMA Journal of Applied Mathematics (London) {\bf 65}, 257 (2000).

\bibitem{80CR}
B. Cotterell and J. R. Rice, Int. J. Fract. {\bf16},   155 (1980).

\bibitem{58Nob}
B. Noble,  {\em Methods Based on the Wiener-Hopf Technique for the solution of Partial Differential
Equations}, (Pergamon, New York, 1958).

  \bibitem{53Mus}
N. I. Muskhelishvili, {\em Some Basic Problems of the Mathematical Theory of
Elasticity}, (Noordhoff, 1953).

  \bibitem{94SSN} S. Sasa, K. Sekimoto and H. Nakanishi, Physical
  Review E \bf{50}\rm, R1733 (1994).


 \bibitem{95bahr} H.A. Bahr, A. Gerbatsch, U. Bahr and H.J. Weiss, Physical
   Review E \bf{52}\rm, 240 (1995).


\bibitem{02LP}
A. Levermann and I. Procaccia, Phys. Rev. Lett., {\bf 89}, 234501 (2002). 

\bibitem{02BLP}
F. Barra, A. Levermann and I. Procaccia, Phys. Rev. E., {\bf 66}, 066122 (2002). 

\end{thebibliography}
\end{document}